\documentclass[twocolumn,astrosymb]{aastex631}
\usepackage{natbib}

\newcommand{\chandra}{{Chandra\/}}

\newcommand{\aspc}{ASP Conf. Ser.}
\newcommand{\spie}{SPIE}
\newcommand{\xray}{{\hbox{X-ray}}}

\usepackage{bm}

\begin{document}
\title{A Rapid and 
Large-Amplitude X-ray Dimming Event in a \boldsymbol{$z\approx2.6$} Radio-Quiet Quasar}

\author{Hezhen Liu}
\affiliation{School of Astronomy and Space Science, Nanjing University, Nanjing, Jiangsu 210093, China}
\affiliation{Department of Astronomy \& Astrophysics, 525 Davey Lab,
The Pennsylvania State University, University Park, PA 16802, USA}
\affiliation{Key Laboratory of Modern Astronomy and Astrophysics (Nanjing University), Ministry of Education, Nanjing 210093, China}

\author{B.~Luo}
\affiliation{School of Astronomy and Space Science, Nanjing University, Nanjing, Jiangsu 210093, China}
\affiliation{Key Laboratory of Modern Astronomy and Astrophysics (Nanjing University), Ministry of Education, Nanjing 210093, China}

\author{W. N. Brandt}
\affiliation{Department of Astronomy \& Astrophysics, 525 Davey Lab,
The Pennsylvania State University, University Park, PA 16802, USA}
\affiliation{Institute for Gravitation and the Cosmos, The Pennsylvania State University,
University Park, PA 16802, USA}
\affiliation{Department of Physics, 104 Davey Lab, The Pennsylvania State University, University Park, PA 16802, USA}

\author{Jian Huang}
\affiliation{School of Astronomy and Space Science, Nanjing University, Nanjing, Jiangsu 210093, China}
\affiliation{Key Laboratory of Modern Astronomy and Astrophysics (Nanjing University), Ministry of Education, Nanjing 210093, China}

\author{Xingting Pu}
\affiliation{College of Science, Nanjing Forestry University, Nanjing, Jiangsu 210037, China}

\author{Weimin Yi}
\affiliation{Yunnan Observatories, Chinese Academy of Sciences, Kunming, 650216, China}
\affiliation{Department of Astronomy \& Astrophysics, 525 Davey Lab,
The Pennsylvania State University, University Park, PA 16802, USA}

\author{Li-Ming Yu}
\affiliation{School of Astronomy and Space Science, Nanjing University, Nanjing, Jiangsu 210093, China}
\affiliation{Key Laboratory of Modern Astronomy and Astrophysics (Nanjing University), Ministry of Education, Nanjing 210093, China}

\begin{abstract}
We report a dramatic fast X-ray dimming event in a $z=2.627$
radio-quiet type 1 quasar,
which has an estimated supermassive black hole (SMBH) mass
of $6.3\times 10^{9} M_\odot$. In the high X-ray state, it showed a typical level of 
X-ray emission relative to its UV/optical emission.
Then its 0.5--2 keV (rest-frame 1.8--7.3 keV) 
flux dropped by a factor of {$\approx7.6$}
within two rest-frame days. 
The dimming is associated with
spectral hardening, as the 2--7 keV (rest-frame 7.3--25.4 keV) 
flux dropped by only $17\%$ and the 
effective power-law photon index of
the \xray\ spectrum changed from $\approx2.3$ to $\approx0.9$.
The quasar has an infrared (IR)-to-UV spectral energy distribution and a 
rest-frame UV spectrum similar to those of typical quasars, and it
does not show any
significant long-term variability in the IR and UV/optical bands.
Such an extremely fast and \hbox{large-amplitude} X-ray variability
event has not been reported before in luminous quasars with such massive
SMBHs. 
The X-ray dimming is best explained by a fast-moving absorber crossing the line of
sight and fully covering the X-ray emitting corona.
Adopting a conservatively small size of {$5  {G} M_{\rm BH}/c^2$}
for the \hbox{X-ray} corona, the transverse velocity of the absorber
is estimated to be {$\approx 0.9c$}. 
The quasar is likely accreting with a high
or even super-Eddington accretion rate, 
and the high-velocity X-ray absorber is probably related to a powerful accretion-disk
wind. Such an energetic wind may eventually
evolve into a massive galactic-scale outflow,
providing efficient feedback to the host galaxy.
\end{abstract}

\section{INTRODUCTION}

Active galactic nuclei (AGNs) are powered by 
accretion onto supermassive black holes (SMBHs) in galaxy centers,
and they are characterized by luminous radiation plus emission variability
across the electromagnetic spectrum. 
AGN multiwavelength variability may involve various physical processes,
and observations of these
have provided useful insights into the still elusive SMBH accretion physics.
With the accumulation of astronomical data over long time spans,
novel types of AGN-related variability phenomena are being discovered;
e.g., the ``changing-look'' quasars \citep[e.g.,][]{Lamassa2015} and the
X-ray quasi-periodic eruptions \citep[e.g.,][]{Miniutti2019,Arcodia2021}. 

Among the multiwavelength variability of \hbox{radio-quiet} AGNs (without strong jets),
\xray\ variability generally occurs on the shortest timescales 
(e.g., \citealt{Ulrich1997}), indicating an origin close to the central SMBH.
AGN X-ray emission is considered to originate from the hot corona surrounding 
the inner accretion disk in the immediate vicinity of the SMBH
\citep[e.g.,][]{Gilfanov2014,Fabian2017}. The typical size of 
the corona is constrained to be around $10r_{\rm g}$, 
where $r_{\rm g}={G} M_{\rm BH}/c^2$ is the gravitational radius with 
$M_{\rm BH}$ being the SMBH mass, 
from studies of rapid \xray\ variability and X-ray microlensing
of AGNs \citep[e.g.,][]{Dai2010,Morgan2012,Shemmer2014,Reis2013,Fabian2015}.
Therefore, the shortest X-ray variability timescale apparently has a SMBH-mass dependence;
it is down to minutes in moderate-luminosity 
AGNs with $M_{\rm BH}\sim10^{6\textrm{--}7} M_\odot$ (e.g., 
IRAS $13224-3809$; \citealt{Fabian2013}), and it is around hours in luminous
quasars with $M_{\rm BH}\gtrsim10^{8} M_\odot$ (e.g., PHL 1092; \citealt{Brandt1999,Miniutti2012}).

Radio-quiet AGNs also vary most strongly at \xray\ energies in general.
The X-ray variability amplitude (here parameterized as the fractional flux change) 
has a broad range. 
The average variability 
amplitude is {$\approx 20\textrm{--}50\%$} for typical AGN samples, and amplitudes
exceeding $\approx2$ are rare
\citep[e.g.,][]{Gibson2012,Yang2016,Maughan2019,Timlin2020}. However, for individual AGNs, 
larger amplitude variability events have been observed with X-ray fluxes changing by
factors of up to a few hundreds. 
For such objects, the X-ray flux may vary strongly with an amplitude of $\approx2$ 
on the shortest timescales mentioned above (minutes to hours),
and the amplitude generally increases with the timescale probed, reaching a
maximum on year timescales.

There are a number of physical interpretations for the broad range of AGN X-ray
variability. The small amplitude variability among typical AGNs is generally
attributed to small fluctuations of the coronal properties (e.g., energy dissipation
from magnetic flares or variation of the optical depth; \citealt{Ulrich1997}).
The strong (amplitudes $\gtrsim2$) or extreme (amplitudes $\gtrsim10$)
variability events are much rarer, and they call for 
significant changes of the coronal properties including, for example, 
the global accretion rate (e.g., changing-look AGNs and tidal disruption events) and
the height of the corona to the SMBH (affecting the relativistic light bending and 
reflection effects; e.g., \citealt{Ross2005,Fabian2012,Dauser2016}).
Alternatively, the strong or extreme variability might not be intrinsic to the corona
but instead be due to obscuration of the coronal X-ray emission
by a varying (e.g., covering factor, column density, or ionization state)
X-ray absorber along the line of sight.
These internal and external mechanisms are not exclusive; in some cases,
changes of the coronal properties and X-ray obscuration are both invoked to
explain the observed X-ray variability \citep[e.g.,][]{Boller2021}.
Due to the often limited quality of AGN X-ray data and the complicated SMBH
accretion physics, we still lack good understanding of the physical processes 
responsible for their X-ray variability.

There is a special type of 
X-ray variability that has attracted much attention 
recently \citep[e.g.,][]{Miniutti2012,Liu2019,Ni2020,Timlin2020,Boller2021,Liu2021}. 
Besides the strong (and often extreme) variability amplitudes, a characteristic property
is that these events are found in type 1 AGNs including luminous quasars and there
is no contemporaneous UV/optical continuum or emission-line variability.
Empirically, there is a significant correlation observed
between the UV/optical emission and X-ray
emission in typical AGNs 
that is interpreted as an intrinsic accretion disk-corona connection
and is often parameterized as a negative relation
between the
monochromatic luminosity at $2500~\textup{\AA}$ ($L_{2500~\textup{\AA}}$)
and the X-ray-to-optical power-law slope
parameter ($\alpha_{\rm OX}$)\footnote{$\alpha_{\rm OX}$ is defined as 
$\alpha_{\rm OX}=0.3838{\rm log}(f_{\rm 2~keV}/
f_{2500~\textup{\AA}})$, where 
$f_{\rm 2keV}$ and $f_{\rm 2500~{\textup{\AA}}}$ are
the \hbox{rest-frame} 2~keV and
2500~\AA\ flux densities.}
or a positive
$L_{2500~\textup{\AA}}$--$L_{\rm 2~keV}$ relation
\citep[e.g.,][]{Strateva2005,Steffen2006,Gibson2008a,Lusso2010,Lusso2016,Risaliti2019,Liu2021}.
Therefore, from the stable UV/optical emission, 
one could infer that the intrinsic coronal X-ray emission 
should not change significantly. Interestingly, the available $\alpha_{\rm OX}$
values computed in the highest X-ray states of these AGNs all follow the
$\alpha_{\rm OX}\textrm{--}L_{2500~\textup{\AA}}$ relation within the scatter 
\citep[e.g.,][]{Miniutti2012,Liu2019,Ni2020,Boller2021,Liu2021},
and they became smaller (more negative; representing \xray\ weakness)\footnote{
In this study, we consider an AGN X-ray weak or in an X-ray weak state if its observed
$\alpha_{\rm OX}$ value deviates significantly below the expectation from the 
$\alpha_{\rm OX}\textrm{--}L_{2500~\textup{\AA}}$ relation. This represents apparent 
weakness of the \xray\ emission, which does not necessarily correspond to 
intrinsic X-ray weakness due to a weak corona.}
in the other 
states. Thus these AGNs likely vary
between the X-ray nominal-strength state and multiple \xray\ weak states, arguing for
the X-ray obscuration scenario where the coronal X-ray emission does not change and the
observed X-ray emission is simply
modified by various amounts of absorption from a small-scale dust-free
\xray\ absorber. These AGNs are generally
considered to have high or even super-Eddington accretion rates, and the \xray\ absorber
is probably related to accretion-disk winds and/or thick inner accretion disks
\citep[e.g.,][]{Miniutti2012,Liu2019,Ni2020,Boller2021}.

In this paper, we report a dramatic fast X-ray dimming event
in a $z=2.627$ \hbox{radio-quiet}
type 1 quasar,
SDSS J$135058.12+261855.2$ (hereafter SDSS J$1350+2618$).
{Its coordinates are $\alpha_{\rm J2000.0}=
13^{\rm h}50^{\rm m}58\fs12$ and $\delta_{\rm J2000.0}=+26\degr18\arcmin55\farcs3$.}
We describe the basic properties of this quasar and the X-ray
data analysis in Section 2. {The exceptional X-ray variability
is presented in Section~3. We show the
spectral energy distribution (SED), 
infrared (IR)-to-UV light curves, and rest-frame UV spectra in Section 4.} 
In Section 5, we discuss physical interpretation of the extreme X-ray variability 
and 
its implications. We summarize in Section 6.
Throughout this paper, we adopt a flat $\Lambda$CDM cosmology with
 the current Planck cosmological parameters \citep{Planck2020} of
$H_{\rm 0}=67.4~\rm km~s^{-1}~Mpc^{-1}$, $\Omega_{\rm M}=0.315$, and 
$\Omega_{\Lambda}=0.685$.

\section{X-ray data analysis}

SDSS J$1350+2618$ is a high-redshift
($z=2.627$; \citealt{Hewett2010}) quasar recently identified as an unusual
quasar that showed significantly weaker \xray\
emission compared to the expectation from the
$\alpha_{\rm OX}\textrm{--}L_{2500~\textup{\AA}}$ relation \citep{Pu2020}.
It is very luminous, with an
absolute $i$-band magnitude (normalized at $z=2$) of $-27.79$, and it
has a {C~{\sc iv}}-based
virial SMBH mass estimate of $6.3\times 10^{9} M_\odot$ \citep{Shen2011}.
Upon detailed investigation, we found that 
this quasar has been serendipitously observed three times by Chandra.
It displayed strong \xray\ variability,
and there was also a state where it showed a nominal level of X-ray emission.
The Chandra observations and the derived \xray\ photometric properties are
listed in Table~1, and the data-analysis procedure is described in the following.

SDSS J$1350+2618$ was serendipitously
observed three times by Chandra ACIS-I during Cycle 18
(observation IDs: 17627, 17621, 17222).
The primary targets of these three \chandra\ observations 
are the outskirts of a galaxy cluster, Abell 1795.
SDSS J$1350+2618$ is $\approx33\arcmin$ away from the cluster center, 
and thus its X-ray data are not contaminated by
the diffuse \xray\ emission of the cluster.
The off-axis angles of SDSS J$1350+2618$ in the three
observations are 6.4\arcmin, 2.6\arcmin, and 6.7\arcmin, respectively.
SDSS J$1350+2618$ landed on the ACIS-I CCD gap in the second observation,
and thus its effective exposure time in this observation is the lowest
among the three
despite the longest total exposure time and smallest off-axis angle.
The CCD gap effects were accounted for by the
effective exposure time and the spectral response files used in the
following photometric and spectroscopic analyses.

We analyzed the Chandra data using the
 Chandra Interactive Analysis of Observations (CIAO; v4.11) tools.
 For each observation,
 we first ran the {\sc chandra\_repro} script to generate a new
 level 2 event file, and then filtered high-background flares 
 using the {\sc deflare} script with an iterative $3\sigma$ clipping
 algorithm; {only $\approx1\%$ of the total exposure was excluded}. 
 From the cleaned level 2 event file, images in the soft
\hbox{(0.5--2~keV)}, hard (2--7~keV), and full (0.5--7~keV) bands were
generated using the {\sc dmcopy} tool. We then ran the automated
source-detection tool {\sc wavdetect} \citep{Freeman2002} to search for
\xray\ sources in the three images, with a false-positive probability
threshold of $10^{-6}$ and wavelet scale sizes of 1, 1.414, 2, 2.828,
4, 5.656, and 8 pixels.
We matched \xray\ positions of the detected sources in each image to
the optical position of SDSS J$1350+2618$.
If the optical position of SDSS J$1350+2618$ was
matched within 2\arcsec\ of an \xray\ position, we considered
SDSS J$1350+2618$ detected by {\sc wavdetect} and
adopted the \xray\ position as the source position to perform
aperture photometry. Otherwise, the optical position was adopted.
For the first observation, SDSS J$1350+2618$ was not detected by
{\sc wavdetect} in any bands.
For the second observation, it was detected in all three
 bands. For the third observation, it was detected in the
 hard and full bands, but not in the soft band.

 The \chandra\ point-spread function (PSF) varies with energy as well
 as off-axis angle, and the off-axis PSF shape is approximately an
 ellipse. In order to minimize the number of background counts
 enclosed by the
 source extraction region and obtain a maximum signal-to-noise ratio, we used
 an elliptical region as the source aperture, which was determined
  with the following procedure.
For each of the three images in each observation,
we simulated an \xray\ source at the SDSS J$1350+2618$ position on the CCD chip, using
 the \chandra\ Ray Tracer (ChaRT; \citealt{Carter2003}) and
 Model of AXAF Response to X-rays (MARX; \citealt{Davis2012})
 software suites. ChaRT was first run to trace rays through the
 \chandra\ \xray\ optics, using the source coordinates,
 an assumed $\Gamma=2$ power-law spectrum, and the observation ID as
 inputs. The collected rays were
 projected to the detector plane via MARX, taking into account all
 detector effects. An image that consists of the simulated photons was
 thus created. The {\sc dmellipse} tool was then used to determine
 an elliptical region that encloses 85\%--94\% of
 the simulated photon flux;
 the elliptical size and the corresponding encircled-energy fraction (EEF)
were chosen
 so that we can obtain the maximum signal-to-noise ratio for the
 source in the real image. The soft- and hard-band images with the
source extraction regions are shown in Figure~1.

\begin{figure*}
 \begin{minipage}[b]{1\linewidth}
\centerline{
\hspace{0.0cm}\raisebox{0pt}{\includegraphics[scale=0.5]{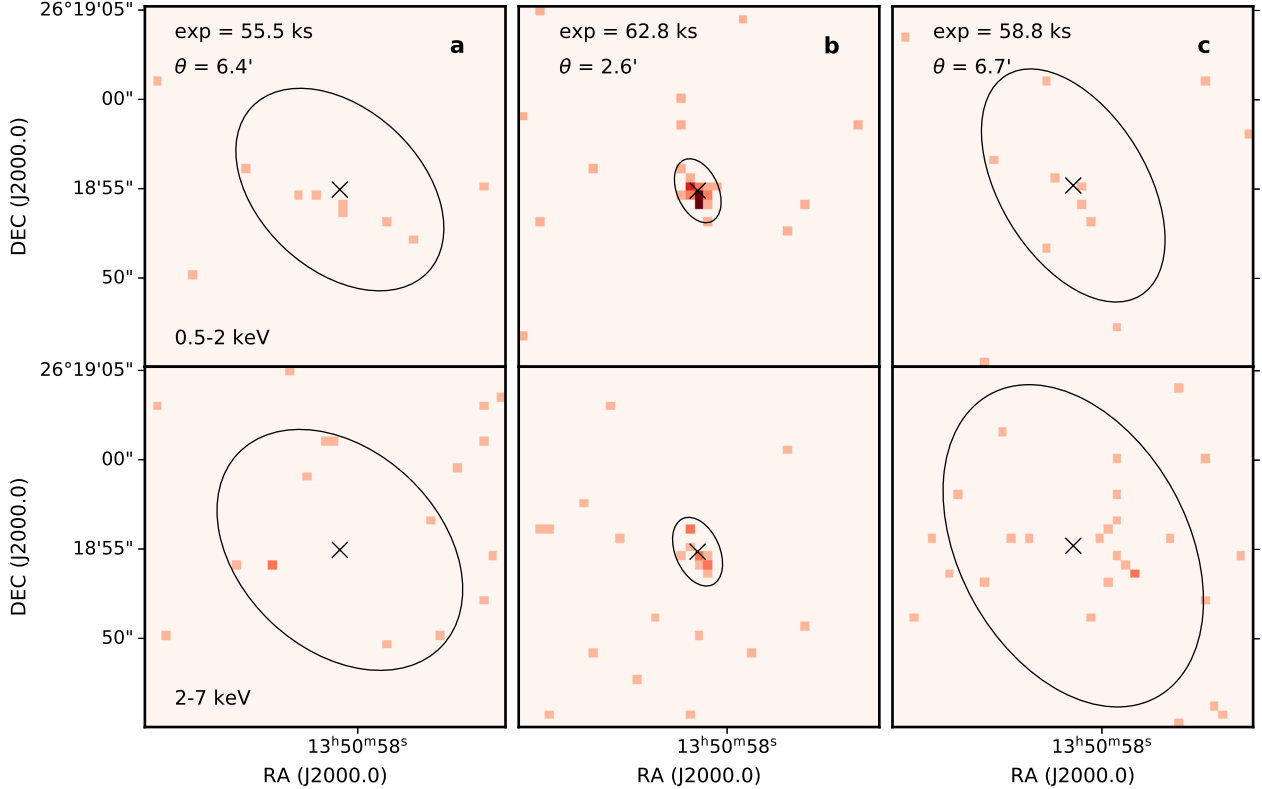}}}
\caption{Chandra 0.5--2~keV (upper panel) and 2--7 keV (lower panel) 
images  of the three observations for 
SDSS J{$1350+2618$}. The observational dates are
{a} 2015 Aug 28, {b} 2016 Mar 27, and {c} 2016 Apr 3.
The ellipse in each image
is the source extraction region used for aperture photometry.
The aperture sizes are chosen so that the signal-to-noise ratios
are the largest.
For observations {a} and {c}, the encircled-energy fractions (EEFs)
of the source apertures are all 94\%; for observation {b},
the EEF is 90\% (85\%) for the soft (hard) band.
The total exposure time and off-axis angle ($\theta$)
of each observation are noted in each of the 0.5--2 keV images.
}
\end{minipage}
\end{figure*}

We extracted source counts ($S$) from
the elliptical region determined above.
Background counts ($B$) were extracted from an annulus region centered
on the source position with a 10\arcsec\ inner radius and a
50\arcsec\ outer radius. We have verified that there is no source detected
by {\sc wavdetect} in any of the background regions.
 We then assessed the source-detection significance via computing the binomial
no-source probability \citep[e.g.,][]{Luo2015,Liu2018}, $P_{\rm B}$, with
the expression:
\begin{equation}
P_{\rm B}=\sum_{X=S}^{N}\frac{N!}{X!(N-X)!}p^X(1-p)^{N-X}~,
\end{equation}
where $N$ is the total number of raw source and background counts ($=S+B$),
and $p=1/(1+BACKSCAL)$ with $BACKSCAL$ being the ratio between the
areas of the background and source regions.
The computed $P_{\rm B}$ values are all
smaller than $0.01$, corresponding to $>2.6\sigma$ detection significance levels.
SDSS J$1350+2618$ is thus considered to be detected in every image for the following photometric analysis.
We calculated $1\sigma$ errors of the source and background counts following the
Poisson approach of \citet{Gehrels1986}, and errors of
the net counts were propagated from the errors of the source and background counts.
The net counts and associated errors in the soft and hard bands are listed in Table~1.

We derived an effective photon index ($\Gamma_{\rm eff}$) for
a power-law spectrum from the observed band ratio, defined as
the ratio of the hard-band to soft-band counts, following a
procedure similar to that in \citet{Liu2018}.
For each observation, we used the {\sc fakeit} routine in XSPEC
(v12.10.1; \citealt{Arnaud1996}) to generate a
set of simulated spectra based on the response matrices and
ancillary response functions in the source-extraction regions of the
soft- and hard-band images, where a Galactic absorption ($N_{\rm H}=1.12\times10^{20}~\rm cm^{-2}$; \citealt{HI4PI2016}) modified
power-law model with a set of $\Gamma$ values was assumed.
We then calculated the corresponding band ratio from the
simulated spectra for each of the assumed $\Gamma$ value. The $\Gamma_{\rm eff}$ values were then derived
from the observed band ratios by interpolating the
$\Gamma$--band ratio sets. The associated $1\sigma$ errors
were propagated from the $1\sigma$
errors of the observed band ratios that were computed using the
Bayesian code BEHR \citep{Park2006}.
For each energy band, we converted the observed
source count rate to flux, with a conversion factor derived from
the simulated spectrum with photon index $\Gamma_{\rm eff}$.
The flux errors were propagated from the errors of the net
counts. Table~1 lists the $\Gamma_{\rm eff}$ and flux values.

 We also extracted the spectrum of the second 
 observation and performed spectral fitting using XSPEC. 
The source spectrum was extracted from
 an elliptical region with an EEF of $90\%$ in the full band,
 and the background spectrum
 was extracted from the same annulus region used
 in the aperture photometry described above.
  The source spectrum was
 grouped with at least one count per bin. We fitted the 0.5--7 keV
 spectrum with a power-law model modified by the Galactic absorption
 ({\sc phabs*zpowerlw}),
 and the W statistic was used in parameter estimation.\footnote{\url{
https://heasarc.gsfc.nasa.gov/docs/xanadu/xspec/manual/XSappendixStatistics.html.}} 
 The best-fit photon index is $2.1\pm 0.4$,
 consistent with the $\Gamma_{\rm eff}$ value within the
 uncertainties. The model-predicted fluxes in the soft and hard bands
 are also consistent
 with those obtained from the photometric analysis.

\section{X-ray variability}

The strong X-ray variability of SDSS J$1350+2618$ is evident from 
the soft-band fluxes listed
in Table 1.
{The 0.5--2 keV 
(rest-frame 1.8--7.3~keV) flux of the second observation is larger 
than that of the first observation by a factor of  
$6.0_{-2.6}^{+4.6}$, and it is larger
than that of the third observation by a factor of
$7.6_{-3.4}^{+6.9}$.
The 1$\sigma$ uncertainties of these variability factors were 
propagated from the flux uncertainties 
following the
method described in Section 1.7.3 of \citet{Lyons1991}.
On the other hand, the flux variability is much weaker
in the hard band (rest-frame 7.3--25.4~keV). The hard-band flux of the
second observation is only higher than those of the first and 
third observations by factors of $1.8_{-0.9}^{+2.0}$ and $1.2_{-0.5}^{+0.9}$, respectively.
Considering the uncertainties, these hard-band fluxes are actually consistent with 
each other.
The significantly different variability amplitudes in the soft and hard bands imply that 
there is variability in the X-ray spectral shape, which is also indicated by the 
$\Gamma_{\rm eff}$ values in Table~1.
SDSS~J$1350+2618$ exhibits a steep spectral shape in the second observation
($\Gamma_{\rm eff}=2.3\pm0.4$), while the spectra appear flatter in the other two
observations, with $\Gamma_{\rm eff}$ values of $1.4^{+0.9}_{-0.7}$ and $0.9^{+0.7}_{-0.6}$, 
respectively.}

We further quantify the X-ray variability using 
the $\alpha_{\rm OX}$ parameter, which allows us to assess the deviations of 
the levels of the observed X-ray emission from the nominal level expected from
the $\alpha_{\rm OX}\textrm{--}L_{2500~\textup{\AA}}$ relation.
We used the \hbox{soft-band}
fluxes to
derive the flux densities at rest-frame 2~keV, assuming a power-law model with the measured
$\Gamma_{\rm eff}$ values. The $2500~\textup{\AA}$ flux density was determined
through interpolation of the photometric data from the Sloan Digital
Sky Survey (SDSS) \citep{York2000}, which is consistent with the value
provided by \citet{Shen2011}. 
{We note that although the X-ray and SDSS observations are not simultaneous,
there does not appear to be any significant UV/optical variability (see Section~4 below)
that could bias the $\alpha_{\rm OX}$ measurements.}
The derived $\alpha_{\rm OX}$ values are listed in
Table~1. We then
calculated the differences ($\Delta\alpha_{\rm OX}$) between the
observed $\alpha_{\rm OX}$ values and those expected from the
$\alpha_{\rm OX}\textrm{--}L_{2500~\textup{\AA}}$ relation in
\citet{Steffen2006}, and the results are also listed in Table~1.
SDSS J$1350+2618$ showed nominal-strength
\xray\ emission in the second observation, with $\Delta\alpha_{\rm OX}=0.03$, while it
showed weak \xray\ emission in the first and third observations, by factors
of
$8.7^{+6.3}_{-3.4}$ and $16.8^{+14.7}_{-6.7}$ ($10^{-\frac{\Delta\alpha_{\rm OX}}{0.3838}}$ 
with the 1$\sigma$ uncertainties
propagated from the flux uncertainties), respectively.
{These X-ray weakness factors (based on the 2~keV flux densities) 
differ from the 0.5--2~keV flux variability amplitudes
computed above ($\approx 6.0$ and $7.6$),
mainly due to 
different $\Gamma_{\rm eff}$
values of the three observations which were used to convert fluxes to flux densities.}

Therefore, SDSS J$1350+2618$ exhibited faint \xray\ emission
in 2015 August, with a rest-frame 2~keV flux density $\approx8.7$
times weaker compared to the expectation from its UV/optical
flux. In 2016 March (two months later in the quasar rest frame),
the second \chandra\ observation revealed that
the quasar recovered to an \xray\ nominal-strength state. The third observation in 2016 April
($47.2$ hours later in the rest frame) surprisingly showed that
the quasar had dimmed by a factor of $\approx 7.6$ in terms of its
0.5--2 keV (rest-frame 1.8--7.3 keV) flux. The dimming is associated with
spectral hardening, as the 2--7~keV (rest-frame 7.3--25.4 keV) flux dropped by only $17\%$.
The effective power-law photon index ($\Gamma_{\rm eff}$) of
the \xray\ spectrum changed from $2.3\pm0.4$ to $0.9_{-0.6}^{+0.7}$.
The most remarkable finding here is the extremely fast 
\xray\ dimming from an X-ray nominal-strength state to a significantly X-ray weak state within
two rest-frame days, which has never been
observed before among luminous quasars with such massive SMBHs.

\section{Spectral energy distribution and multiwavelength properties}
We explored if SDSS J$1350+2618$ displays
any unusual features in
its multiwavelength data.
We collected
IR-to-UV photometric data for SDSS J$1350+2618$, in order
to construct its SED and explore if
there is notable variability in the IR-to-UV bands.
The data were collected from the public catalogs of the
Wide-field~Infrared~Survey~Explorer (WISE; \citealt{Wright2010}),
Near-Earth Object WISE Reactivation
(NEOWISE; \citealt{Mainzer2014}),
UKIRT Infrared Deep Sky Survey (UKIDSS; \citealt{Lawrence2007}),
Pan-STARRS1 Surveys (PS1; \citealt{Chambers2016}),
SDSS, Zwicky Transient Facility (ZTF; \citealt{Bellm2019}), and
Catalina Real-Time Transient Survey (CRTS; \citealt{Drake2009}).
All the data were corrected for Galactic extinction
($E_{\rm B-V}=0.0122$; \citealt{Schlegel1998})
using the extinction law of \citet{Cardelli1989}.
Except for the UKIDSS and SDSS catalogs that provide only single-epoch measurements
and the WISE catalog that has stacked measurements,
the other surveys provide multiple measurements from long-term
multi-epoch observations, and we used their error-weighted average
measurements to construct the SED, shown in Figure~2. Particularly, for the
 NEOWISE data, we plotted the average measurements in
 the day closest to the third \chandra\ observation.
For the X-ray SED, we derived
the 2 keV and 10 keV luminosities from the
0.5--2 keV and 2--7 keV fluxes, adopting a
power-law model with the $\Gamma_{\rm eff}$ values from the aperture
photometry.
For comparison, the
mean SED of
SDSS quasars in \citet{Krawczyk2013} is also shown, which is
scaled to the mean $2500~{\textup{\AA}}$ luminosity of SDSS~J$1350+2618$ interpolated from
the SDSS photometric data.
Except for the stronger mid-IR emission,
SDSS~J$1350+2618$ shows a fairly typical quasar IR-to-UV SED.
{The strong X-ray variability is clearly visible in the SED plot, which also indicates that
SDSS~J$1350+2618$ was in
an X-ray nominal-strength state during the second observation with a steep spectral shape, and 
it switched to 
X-ray weak states in the first and 
third observations with flatter spectral shapes.}

\begin{figure*}
 \begin{minipage}[b]{1\linewidth}
\centerline{
\hspace{0.0cm}\raisebox{0pt}{\includegraphics[scale=0.6]{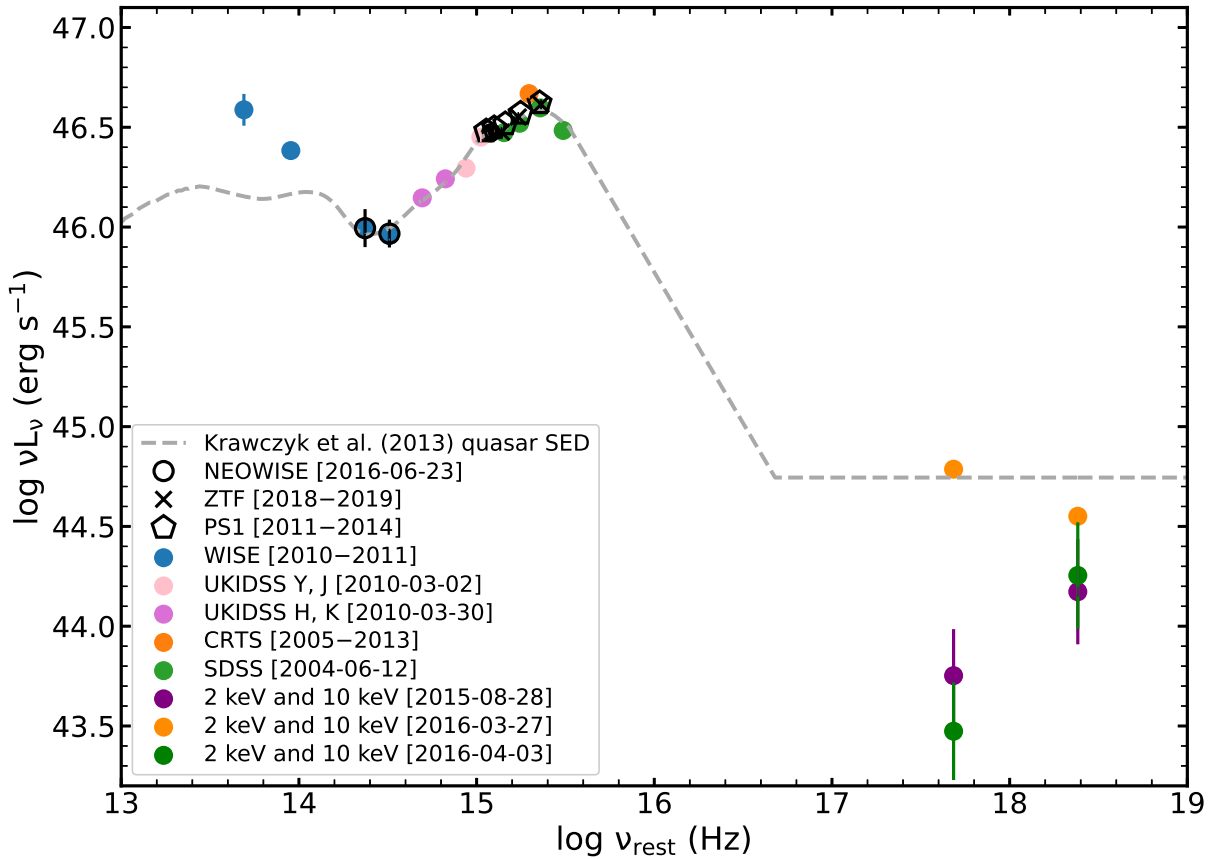}}}
\caption{IR-to-X-ray SED for SDSS~J{$1350+2618$}. The IR-to-UV
data points were gathered from the WISE, NEOWISE, UKIDSS,
PS1, SDSS, ZTF, and CRTS catalogs, where the WISE, PS1, ZTF,
and CRTS
data points show the average measurements of multi-epoch
observations. The rest-frame 2 keV and 10 keV luminosities for the
three \chandra\ observations were derived from the
observed 0.5--2 keV and 2--7 keV fluxes, adopting a
power-law model with the measured $\Gamma_{\rm eff}$ values.
The corresponding observational dates/periods are listed in the lower
left corner. The data error bars show
$1\sigma$ uncertainties, and some of them are
smaller than the symbol size.
The dashed curve shows the mean SED of
SDSS quasars in \citet{Krawczyk2013}, which is
scaled to the $2500~{\textup{\AA}}$ luminosity of SDSS~J$1350+2618$ interpolated from
the SDSS photometric data. Except for the stronger mid-IR emission,
SDSS~J$1350+2618$ shows an IR-to-UV SED consistent with the mean quasar
SED.
}
\end{minipage}
\end{figure*}

Through integrating the scaled $1\rm~\mu m\textrm{--}
10$~keV mean quasar SED shown in Figure~2, we estimate a bolometric
luminosity ($L_{\rm Bol}$) of $1.3\times10^{47}$~erg~s$^{-1}$
for SDSS J$1350+2618$.
With the Eddington luminosity computed as $L_{\rm Edd} = 1.26\times10^{38} M_{\rm BH}$~erg~s$^{-1}$,
the Eddington ratio ($L_{\rm Bol}/L_{\rm Edd}$) of this
quasar is $\approx0.16$, consistent with the value provided in \citet{Shen2011}.
This Eddington ratio is within the interquartile range of $\approx0.15\textrm{--}0.46$ for the
Eddington ratios of typical quasars at $z=2\textrm{--} 3$ (\citealt{Shen2011}), suggesting that the SMBH
mass estimate for SDSS J$1350+2618$, although uncertain, is not off by a large factor.
Its large X-ray photon index of $2.3\pm0.4$ in the X-ray nominal-strength state
also suggests that SDSS J$1350+2618$ is accreting with
a high or even super-Eddington accretion rate \citep[e.g.,][]{Shemmer2008}.

{We then constructed 
IR-to-UV light curves obtained from
the PS1, ZTF, CRTS, and NEOWISE catalogs, shown in Figure~3.}
For the ZTF, CRTS, and NEOWISE light curves,
we grouped any intra-day measurements.
{In the PS1 light curves, we also included the SDSS $g$-, $r$-, $i$-, $z$-band 
photometric measurements 
and the $g$-, $r$-, $i$-band magnitudes
derived from the SDSS spectrum.
In the CRTS light curve, we added the $V$-band magnitude
derived from the SDSS spectrum, of which the observational date overlaps the CRTS light curve.
These light curves indicate that SDSS~J$1350+2618$ does not show any
substantial long-term variability in the IR and UV/optical bands.
Although only the NEOWISE light curve overlaps the \xray\ observational dates,
the mild variability with amplitudes $\lesssim1.5$ ($\approx0.4$ mag) 
in all the rest-frame IR-to-UV bands shown in Figure~3 indicates that there were unlikely
any significant changes of the accretion power. Thus, the $\alpha_{\rm OX}$ values and \xray\ states (nominal-strength state for the second observation
and weak states for the first and
third observations)
of SDSS~J$1350+2618$, determined using the non-simultaneous X-ray and
SDSS photometric (for $f_{2500~\textup{\AA}}$ interpolation) data, should not be 
affected by any potential strong UV variability.
}

\begin{figure*}
\centering
\includegraphics[trim=0 10 20 20,clip, width=0.49\linewidth]{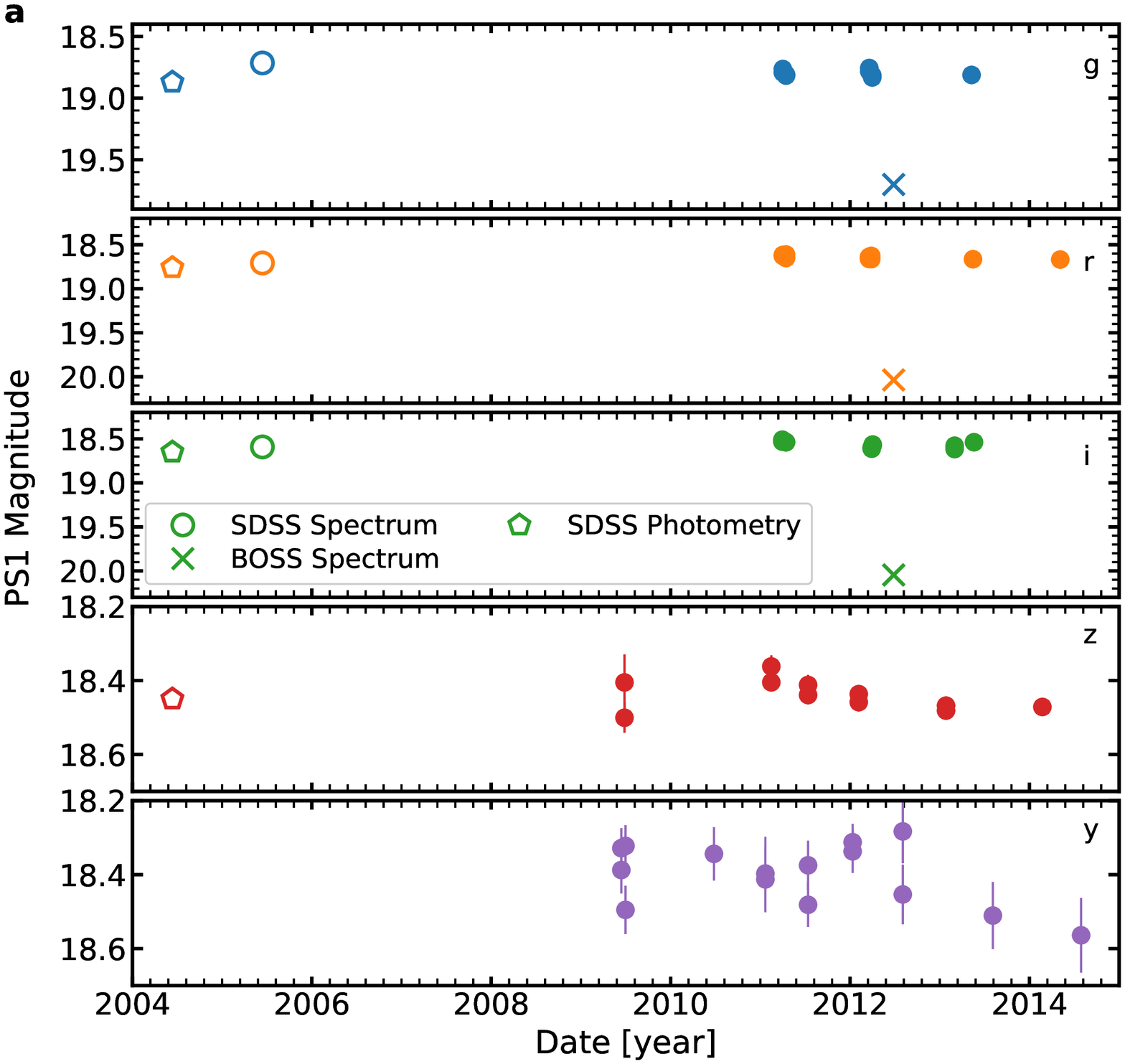}
\includegraphics[trim=0 10 20 20,clip, width=0.49\linewidth]{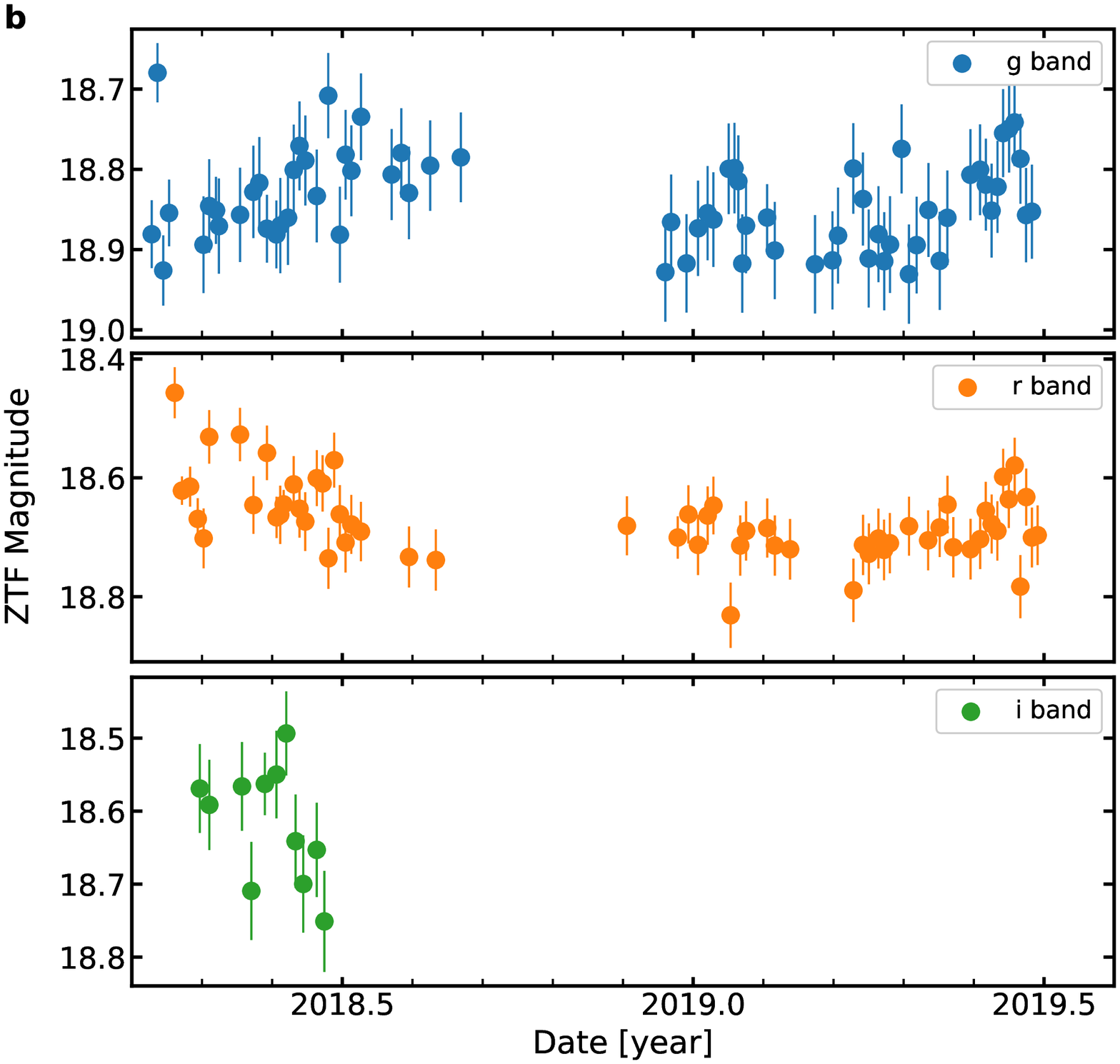}
\includegraphics[trim=0 10 20 20,clip,width=0.49\linewidth]{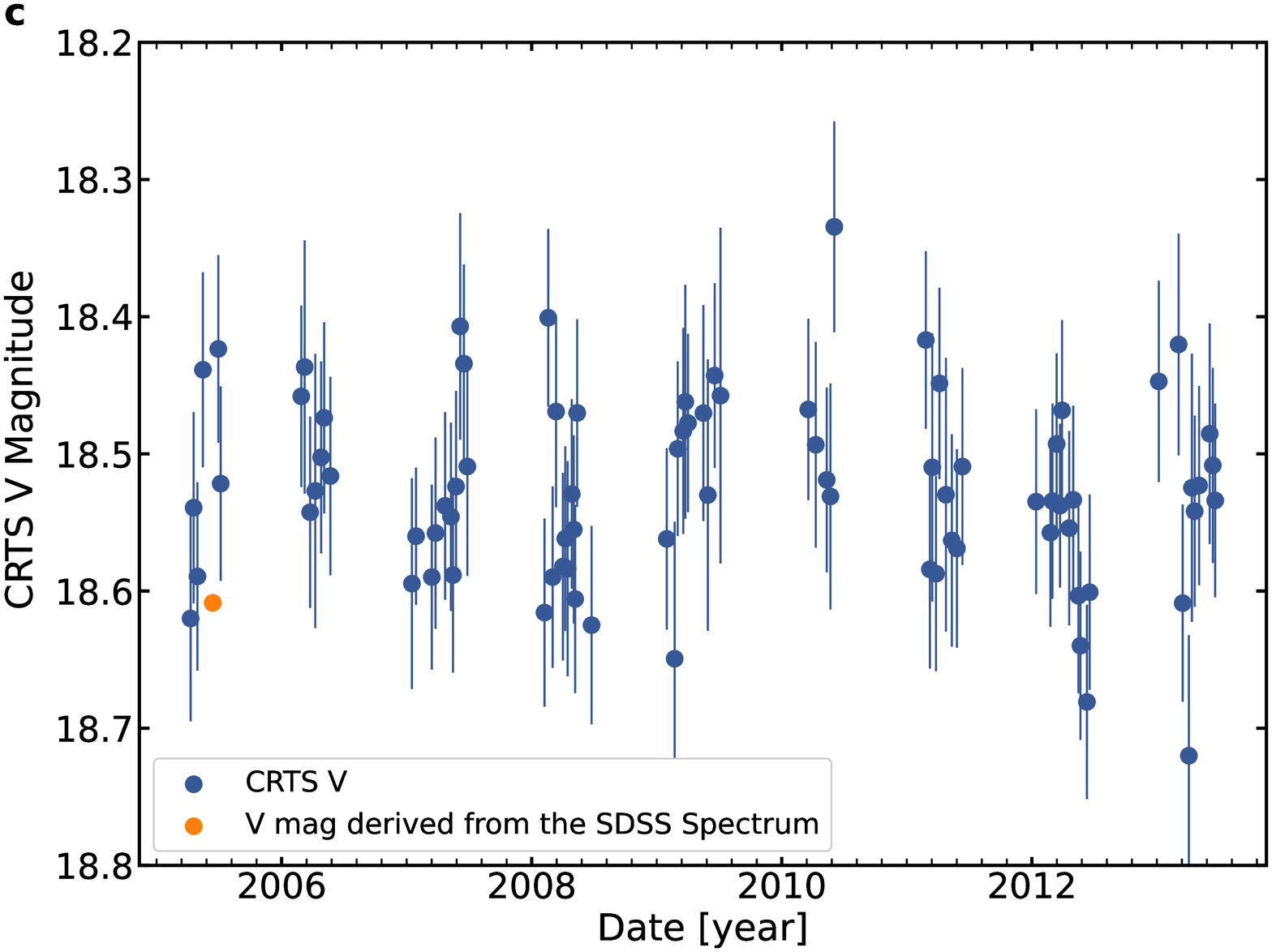}
\includegraphics[trim=0 10 20 20,clip,width=0.49\linewidth]{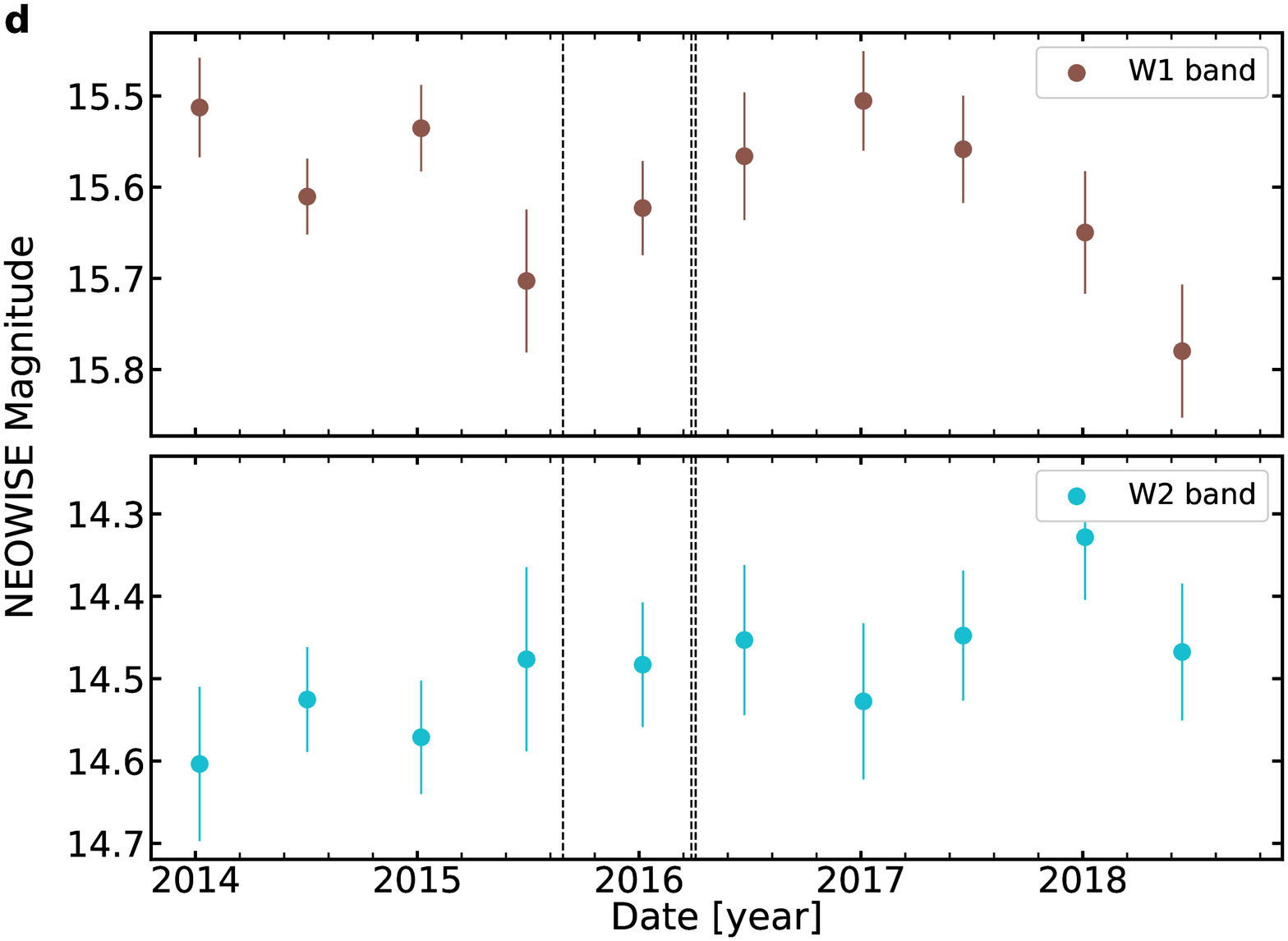}
\caption{IR-to-UV light curves.
{a}, PS1 $g$, $r$, $i$, $z$, and $y$ bands.
{b}, ZTF $g$, $r$, and $i$ bands.
{c}, CRTS $V$ band.
{d}, NEOWISE W1 and W2 bands.
For the ZTF, CRTS, and NEOWISE light curves,
we grouped any intra-day measurements.
{
In the PS1 light curves, we also included the SDSS $g$-, $r$-, $i$-, $z$-band
photometric measurements
and the $g$-, $r$-, $i$-band magnitudes
derived from the SDSS and BOSS spectra.
The BOSS measurements are likely
incorrect, probably due to absolute flux calibration
issues.
In the CRTS light curve, we added the $V$-band magnitude
derived from the SDSS spectrum. In the NEOWISE light curve, the three vertical dashed lines
indicate the observational dates of the X-ray observations.}
These light curves indicate that SDSS~J$1350+2618$ does not show any
substantial long-term variability in the IR and UV/optical bands.
}
\end{figure*}

The SDSS spectrum taken in 2005 is displayed in Figure~4.
The spectrum has been corrected for
the Galactic extinction.
For comparison,
the composite spectrum of SDSS quasars \citep{Vanden2001}
was plotted with a gray curve. The
continuum appears to be slightly bluer than
the composite spectrum.
The {C~{\sc iv}} $\lambda1549$ and {C~{\sc iii}]}~$\lambda1909$ emission lines
are weaker than those in the composite spectrum.
The {C~{\sc iv}} rest-frame equivalent width is only 12.2~\AA\ (\citealt{Shen2011}),
compared to the 30~\AA\ value for the composite spectrum, and thus
SDSS J$1350+2618$ could be classified as a weak emission-line quasar (WLQ),
a small population of type 1 quasars that show unusually weak UV emission
lines.
Similar to many other WLQs, SDSS J$1350+2618$ also shows a large {C~{\sc iv}} blueshift;
the measurement from \citet{Shen2011} adjusted to the improved redshift \citep{Hewett2010} is
2630~km~s$^{-1}$. The prominent {C~{\sc iv}} blueshifts in WLQs suggest
a strong wind component for the {C~{\sc iv}} broad emission-line region (BELR).
There are no
broad absorption features in the rest-frame UV spectrum. But there appear to be a
{C~{\sc iv}} narrow absorption line (NAL) at 1413~\AA\ and a
Ly$\alpha$ NAL at 1177~\AA. 

Besides the SDSS spectrum, there is also a SDSS Baryon Oscillation Spectroscopic Survey (BOSS)
spectrum for SDSS J$1350+2618$ taken in 2012. {The BOSS spectrum, shown in Figure~4,
is very similar to the SDSS spectrum with a similarly weak {C~{\sc iv}} emission line and two similar
NALs, but the
overall flux is lower by a factor of $\approx3$.
The ratio curve 
of the two spectra (Figure~4) is featureless, indicating no apparent spectral variability besides
the flux difference.
The $g$-, $r$-, $i$-band magnitudes
derived from the BOSS spectrum were also presented in the PS1 light curves (Figure~3a) for comparison.}
Since there is no significant variability observed
in the optical photometric light curves
which cover the observational dates of the SDSS and BOSS spectra, we consider the BOSS flux 
measurement
incorrect, probably due to some BOSS absolute flux calibration
issues.\footnote{\url{http://www.sdss3.org/dr9/spectro/caveats.php\#qsoflux.}}

\begin{figure*}
\begin{minipage}[b]{1\linewidth}
\centerline{
\hspace{0.0cm}\raisebox{0pt}{\includegraphics[scale=0.6]{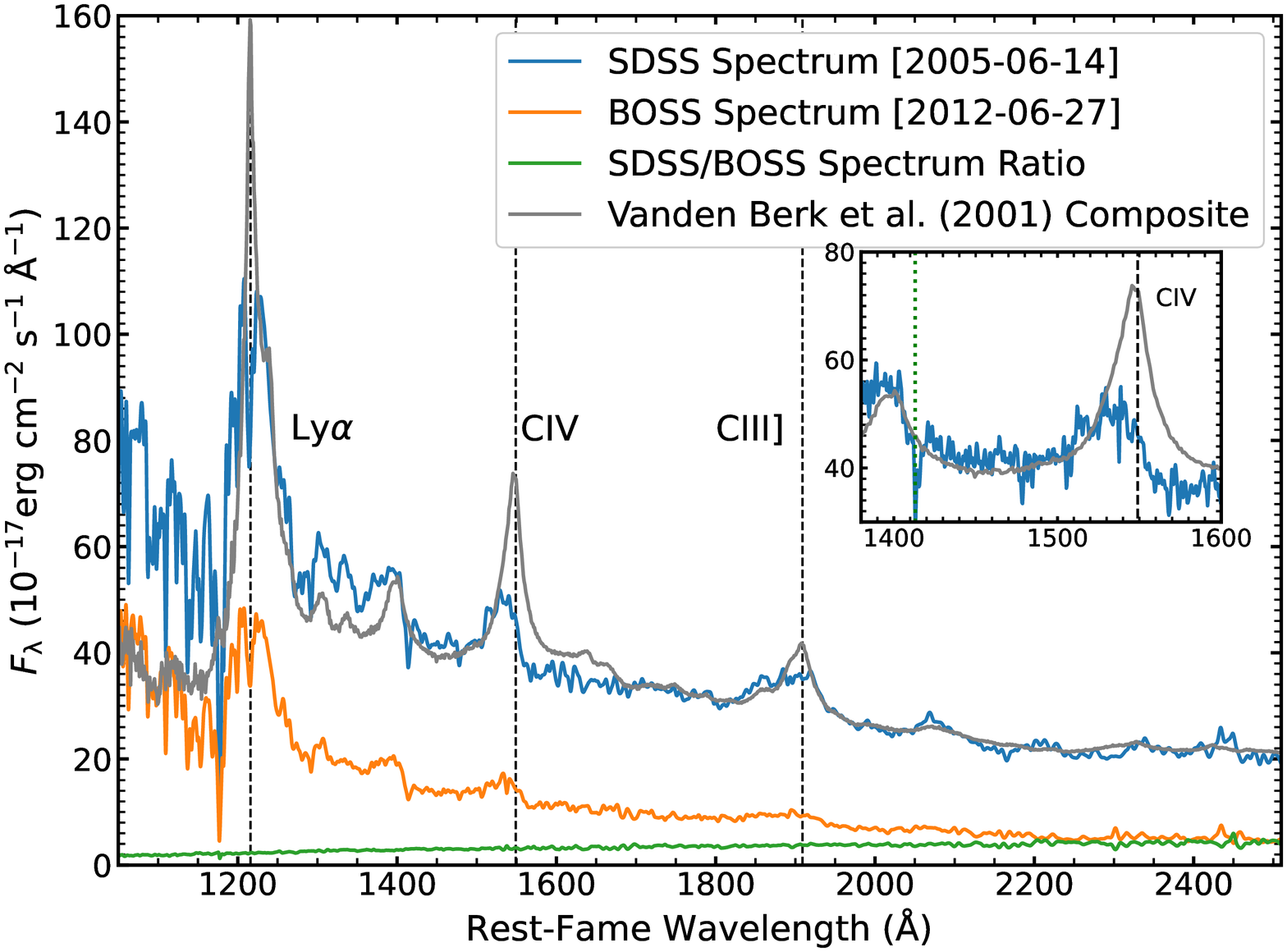}}}
\caption{SDSS spectrum of SDSS J{$1350+2618$} (blue curve).
The inset shows a zoomed-in view of the {C~{\sc iv}} $\lambda1549$ emission line and an
associated NAL at 1413~\AA.
For comparison, the
gray curve shows the composite spectrum of typical SDSS
quasars \citep{Vanden2001}, scaled to the spectrum of SDSS J$1350+2618$ at
rest-frame $1695~\textup{\AA}$.
There are no broad absorption features in the spectrum of
SDSS J$1350+2618$. Compared to typical quasars, SDSS J$1350+2618$
shows a weaker and blueshifted {C~{\sc iv}} emission line.
{The BOSS spectrum is also displayed (orange curve), showing similar spectral features. 
The overall flux of the SDSS spectrum is higher than that of the BOSS spectrum by 
a factor of $\approx3$, and
the green curve shows the ratio of these two spectra. 
We consider the BOSS flux measurement
unreliable, probably due to absolute flux calibration
issues.}
}
\end{minipage}
\end{figure*}

\section{Discussion}
{We presented the extreme X-ray variability of SDSS~J$1350+2618$.
The rapid and large-amplitude X-ray dimming from the second observation to the third observation
appears unique among such luminous quasars.
Unfortunately, the limited photon statistics of the \xray\ data in the third observation
(with $\approx17$ \hbox{0.5--7~keV} net counts)
do not allow detailed spectral modeling; we are not even able to demonstrate conclusively 
that
absorption must be present given the flat but rather uncertain effective photon index
($0.9^{+0.7}_{-0.6}$), although the data suggest this is a likely possibility.
Therefore, the nature of the dimming is 
not immediately clear. In the following discussion, we consider that
the dimming is most naturally explained by an X-ray eclipse event, based on the X-ray 
and multiwavelength properties presented above and our current 
understanding of AGNs with similar properties, although their SMBH masses are not as large
or their variability timescales are not as short.
However, we do not consider this obscuration scenario conclusive. Future observations of
SDSS~J$1350+2618$ or discoveries of similar events should 
shed additional light on the elusive nature.}

\subsection{A Near-Light-Speed X-ray Eclipse Event}

The X-ray variability of SDSS~J$1350+2618$ has a few distinctive features, 
namely, an extremely short timescale of $\approx$ 2 days for a 
$M_{\rm BH}\approx6.3\times 10^{9} M_\odot$ SMBH, variation between X-ray normal
and significantly X-ray weak states, and a typical photon index of 
$\approx2.3$ in the 
\xray\ nominal-strength state and spectral hardening in the \xray\ weak states.
These features immediately suggest varying X-ray obscuration as the most natural 
physical interpretation for the variability. The short variability timescale 
does not permit any significant changes to the accretion rate, 
and thus the scenarios of changing-look events or TDEs can
be excluded; the typical quasar IR-to-UV SED and the lack of 
significant IR-to-UV long-term
variability also argue against these
scenarios.
Besides X-ray obscuration, the other 
scenario that can modify the observed X-ray emission is the relativistic
disk-reflection model \citep[e.g.,][]{Ross2005,Fabian2012,Dauser2016}.
In this case,
the weak \xray\ emission is due to the light bending effects when the corona moves very
close to the SMBH, and the observed spectrum is dominated by a disk-reflected component;
the \xray\ variability is due to changes of the corona height/size which affect the
strengths of the light bending and disk-reflection effects.
However, for SDSS~J$1350+2618$,
the hard-band (rest-frame \hbox{7.3--25.4~keV})
flux in the third observation dropped by only
17\% compared to the \xray\ nominal-strength state
in the second observation that has a typical unobscured power-law spectrum;
this cannot be produced by a disk-reflected spectrum which should be much weaker
than the intrinsic spectrum in the hard band.

Therefore, a change of X-ray obscuration remains the only viable scenario. 
Indeed, the changes of
X-ray fluxes and
spectral shape from the second to the third observation
are all consistent with change from
an unobscured spectrum to a heavily obscured spectrum.
A schematic illustration of the
obscuration scenario in displayed in Figure~5.
Under this scenario, the transverse velocity of the absorber can be
constrained.
Given the occultation time of $47.2$ hours between the second and third observations,
and adopting a conservatively small size of $5r_{\rm g}$
for the \xray\
corona,
the velocity
of the absorber reaches a relativistic value of
$\approx 0.9c$.
Uncertainty of this unprecedentedly high velocity constraint mainly comes from the
uncertain SMBH mass estimate, as there is no reliable method for
measuring SMBH masses of high-redshift quasars.
Nevertheless, the extreme X-ray flux and spectral-shape variability of SDSS J$1350+2618$
suggest a novel eclipse event in the
vicinity of a massive SMBH.

\begin{figure*}
\begin{minipage}[b]{1\linewidth}
\centerline{
\hspace{0.0cm}\raisebox{0pt}{\includegraphics[scale=0.8]{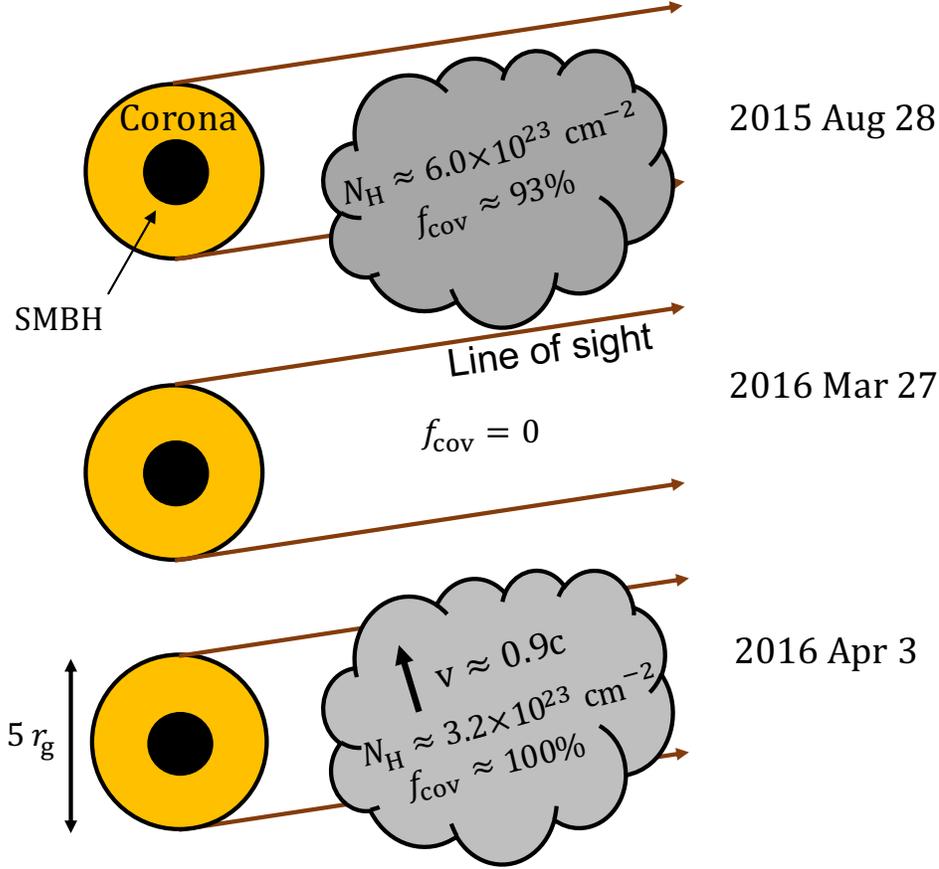}}}
\caption{A schematic diagram of the varying \xray\ obscuration in
SDSS J$1350+2618$.
The moving gas clumps with different sizes and column densities sometimes
cross the line of sight and partially shield
the \xray\ corona. During the first \xray\ observation (2015 Aug 28),
the corona of
SDSS J$1350+2618$ was partially covered by a clump with
a covering factor of $\approx93\%$ and a column density of $\approx6.0\times 10^{23}\ \rm cm^{-2}$.
The corona was not covered by any clumps in the
second observation (2016 Mar 27). Within two rest-frame days,
a fast-moving clump with a column density of
$\approx3.2\times10^{23}\ \rm cm^{-2}$
crossed the line of sight to the corona,
and it fully ($\approx100\%$) covered the corona in the third observation.
The \xray\ absorber is probably not a
large cloud as shown in the cartoon but instead a combination of small gas clumps.
{The column densities of the clumps are estimated from our assumed model and are thus
model dependent (Section 5.1).}
Given the crossing time of $47.2$ hours
and assuming a corona
size of $5r_{\rm g}$ (corresponding to the minimum transverse traveling distance of the clump),
the transverse velocity of this clump is estimated to be
$\approx 0.9c$.
{Uncertainty of this unprecedentedly high velocity constraint mainly comes from the
uncertain SMBH mass estimate.}
}
\end{minipage}
\end{figure*}

We then tried to constrain the properties of the \xray\ absorber by fitting
the X-ray spectra in the first and third observations with a partial-covering absorption model.
{We note that this approach was not to analyze the X-ray data, which was 
presented in Section~2. Instead, we were simply trying to derive
basic physical constraints under our inferred obscuration scenario. The spectra
in the first and third observations are not of 
high quality for complex spectral modeling, and thus we had to fix most of the spectral
parameters at assumed/inferred values.}
The extraction and simple power-law fitting of the spectrum in the
second observation are described in Section 2.
For the first and third observations,
we extracted the source spectra from
elliptical regions with a $95\%$ EEF in the full band and
background spectra from an annulus region with an inner (outer) radius of 10\arcsec\ (50\arcsec).
The spectra were grouped with a
 minimum number of one count per bin, and the W statistic was used in
 parameter estimation.
The overall partial-covering absorption model in XSPEC is
{\sc phabs*(constant1*zpowerlw*zphabs*cabs+
constant2*zpowerlw)},
 where {\sc phabs} is to account for the Galactic absorption,
 {\sc constant1} is the covering factor of the intrinsic absorber, and
{\sc constant2=1-constant1}. The {\sc cabs} component takes into
account the Compton scattering of the absorber with its
column density tied to that of the {\sc zphabs} component.
We
fixed the intrinsic continuum (photon index and normalization of
{\sc zpowerlw}) to the best-fit power law of the second observation.
We further fixed the covering factor in the third observation to $1.0$, as
the flat spectral shape ($\Gamma_{\rm eff}\approx0.9$) does not suggest a partial-covering
absorber.
The best-fit results are shown in Figure~6. For the first observation,
the column density is $(6.0_{-2.3}^{+6.9})\times 10^{23}\ \rm cm^{-2}$ and the covering
factor is $0.93_{-0.12}^{+0.07}$; for the third observation,
the column density is $(3.2_{-1.0}^{+1.7})\times 10^{23}\ \rm cm^{-2}$.
{The quoted errors are at a 68\% ($1\sigma$) confidence level for one parameter of interest.}
These basic absorber properties are included in the schematic illustration of the
obscuration scenario in Figure~5.
We caution that these parameters are derived based on our assumed model
and are thus model dependent. The real absorber might be more complex
(e.g., partially ionized), but the data do not allow constraints on
more complicated models. Our previous argument about the absorber velocity
does not rely on these absorber parameters.

\begin{figure*}
\begin{minipage}[b]{1\linewidth}
\centerline{
\hspace{0.0cm}\raisebox{0pt}{\includegraphics[scale=0.7]{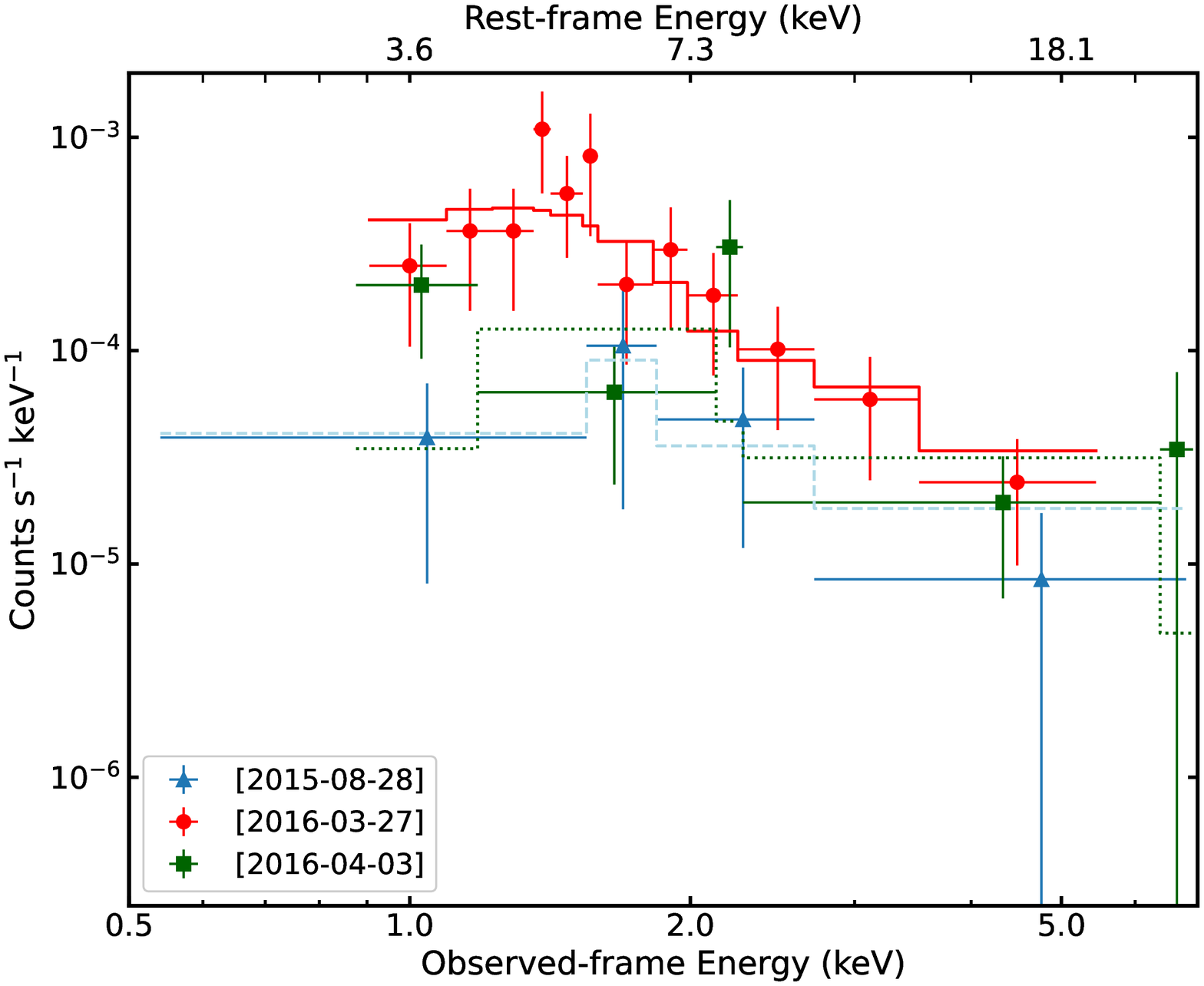}}}
\caption{{Three \chandra\ spectra overlaid with the best-fit models in the obscuration scenario.}
The spectra are grouped for display purposes.
The spectrum of the second observation is fitted with a simple power-law model,
and the best-fit spectrum is fixed as the intrinsic continuum in the
partial-covering absorption modeling of the first and third observations.
Basic absorber properties in the first and third observations are thus derived, as shown
in Figure~5. {We note that this approach was not to analyze the X-ray data, but
to derive basic absorber parameters,
and the results are strongly model dependent. 
The real absorber might be more complex
(e.g., partially ionized), but the data do not allow constraints on
more complicated models. Our previous argument about the absorber velocity
does not rely on these absorber parameters.}
}
\end{minipage}
\end{figure*}

\subsection{Connections to Ultra-Fast Outflows and AGN Feedback}

The high-velocity \xray\ absorber in SDSS J$1350+2618$
is reminiscent of ultra-fast outflows (UFOs) with velocities up to
$0.1\textrm{--}0.4c$ found in an increasing number of low-redshift AGNs 
\citep[e.g.,][]{Tombesi2010,Gofford2013}.
The velocities were inferred from substantially blueshifted iron absorption lines
detected in the $\gtrsim7$ keV X-ray spectra.
The UFOs are considered to be
launched
from the inner accretion disk since the outflow velocities are
comparable to the escape velocities at small radii.
The UFOs are probably clumpy, and the high-density part could provide a large amount of absorption to
the \xray\ continuum and thus be responsible for \hbox{large-amplitude}
\xray\ spectral variability observed in a few local
AGNs \citep[e.g.,][]{Hagino2016,Matzeu2016,Park2018}. Moreover, these UFOs
may carry a large amount of mechanical power, being a
possible source for AGN feedback into the host
galaxy \citep[e.g.,][]{Crenshaw2012,King2015}.
Although UFOs are becoming fairly well characterized in local AGNs, there is still a
lack of understanding about them in the distant universe.
Since high-quality \xray\ spectra are required to diagnose the
presence of hard \xray\ absorption lines, evidence of UFOs has been
collected for only a limited number of high-redshift
 quasars \citep[e.g.,][]{Chartas2002,Chartas2003,Chartas2007,
 Chartas2016,Chartas2021,Dadina2018}.

The \xray\ absorber in SDSS J$1350+2618$ may share the same origin
as the UFOs. Due to the limited photon statistics of
the three \chandra\ spectra, we cannot resolve any \xray\ absorption
lines or determine the ionization state of the \xray\ absorber.
It has been suggested that UV NALs are common in quasars with X-ray identified UFOs and
these outflows are probably linked \citep[e.g.,][]{Chartas2021}.
If the {C~{\sc iv}} and Ly$\alpha$ NALs of SDSS J$1350+2618$ (Section 4) are intrinsic to 
the quasar instead of from
intervening systems, the corresponding outflow velocities are $\approx0.09c$ and $0.03c$, respectively.
Thus, these two NALs might provide
additional support to the presence of the high-velocity X-ray absorber in SDSS J$1350+2618$.
The partial-covering absorber that shielded the \xray\ corona
could be associated with denser and
less-ionized clumps embedded in the inhomogeneous
wind, such as those suggested in \citet{Hagino2016} and \citet{Matzeu2016}.
Clumpy structures are generally expected in
quasar outflows due to thermal
instabilities \citep[e.g.,][]{Waters2017,Dannen2020} and/or radiation hydrodynamic
instabilities \citep[e.g.,][]{Takeuchi2014}.
The stochastic clumps with different sizes and densities in the fast
wind move across our line of sight occasionally, resulting in the observed
large-amplitude fast \xray\ variability.
The $\approx 0.9c$ velocity constrained above for the absorber could be related to the outflow and rotational
velocities of the wind.

Under the clumpy wind scenario,
the relativistic outflow in 
SDSS {J$1350+2618$} may carry a large amount of energy and mass into
the host galaxy, providing efficient AGN feedback. With
the estimated properties of the \xray\ absorber,
the mass outflow rate can be estimated following the
expression of $\dot{M}_{\rm out}=4\pi\mu r N_{\rm H}v_{\rm out}
m_{\rm p}C_{g}$ (\citealt{Crenshaw2012}), where $N_{\rm H}$ and
$v_{\rm out}$ are the column density and velocity of the absorber,
$r$ is the absorber's radial location,
$\mu$ is the mean atomic mass
per particle ($=1.4$ for solar abundances), $m_{\rm P}$ is the proton
mass, and $C_{g}$ is the global covering
fraction of the outflow (assumed to be 0.5 based on the UFO absorber statistics;
\citealt{Crenshaw2012} and references therein).
We assume that
$v_{\rm out}$ equals the transverse velocity ($\approx 0.9c$) of the absorber,
$N_{\rm H}=3.2\times 10^{23}~\rm cm^{-2}$,
and $r=10r_{\rm g}$ considering the high velocity of the absorber.
Under these assumptions, we estimate the mass-outflow
rate to be $\approx 19~M_\odot~{\rm yr^{-1}}$. The energy outflow rate (mechanical power) can thus
be estimated to be $\dot{E}_{\rm out}=\frac{1}{2}\dot{M}_{\rm out}
v_{\rm out}^2\approx 4.3\times 10^{47}~\rm erg~s^{-1}$. This
value exceeds the bolometric luminosity ($1.3\times10^{47}~\rm
erg~\rm s^{-1}$) estimated above, and it is around $54\%$ of the
Eddington luminosity of SDSS J$1350+2618$.
The main uncertainty of this $\dot{E}_{\rm out}$ estimate is 
from the $v_{\rm out}$ (and thus SMBH mass) uncertainty.
Moreover, if the quasar is indeed \hbox{super-Eddington} accreting,
the bolometric luminosity or the Eddington luminosity is not an accurate representative of the
accretion power, as a large
fraction of the energy may be advected into the SMBH or be converted into the mechanical
power of a wind \citep[e.g.,][]{Jiang2019}.
Nevertheless, the outflow efficiency ($\dot{E}_{\rm out}/L_{\rm Bol}$) likely
exceeds the threshold of
$\dot{E}_{\rm out}/L_{\rm Bol}>0.5\textrm{--}5\%$
 for significant feedback into the host
 galaxy \citep[e.g.,][]{Matteo2005,Hopkins2010}. Such an outflow
 has enough power to sweep out dust and gas from the host
 galaxy, quenching the star formation and further
SMBH growth.

\subsection{Connections to Super-Eddington Accreting AGNs and WLQs}

The main observational properties of SDSS J$1350+2618$,
the variability between X-ray normal and \xray\ weak states with no significant
UV/optical variability, have already been found in a small population of AGNs including
high-redshift quasars that are considered to have super-Eddington accretion
rates (Section~1),
although previously discovered variability timescales
are not sufficiently short to constrain a relativistic outflow as in
SDSS J$1350+2618$. 
Powerful radiatively driven accretion-disk winds are
expected in these systems \citep[e.g.,][]{Takeuchi2014,Giustini2019,Jiang2019},
and the denser and
less-ionized clumps of the winds likely provide the varying partial-covering
absorption for interpreting the X-ray variability.
The fraction of super-Eddington accreting AGNs showing such strong X-ray variability 
is poorly constrained ($\sim15\%$; \citealt{Liu2019}), mainly due to the limited numbers
of multi-epoch observations currently available. Inclination angle is probably
a key factor determining whether we can observe (variable) X-ray absorption
in these AGNs. Another important factor might be the Eddington ratio, which likely 
affects the strength/density and covering factor of the clumpy disk wind. 

Similar to SDSS J$1350+2618$, some of these
quasars are classified as WLQs with unusually weak 
UV high-ionization emission lines (e.g., {C~{\sc iv}}; Figure 4).
Systematic studies of WLQ X-ray emission have revealed that
a large fraction ($\sim50\%$) of them are X-ray weak while
the others show typical levels of X-ray emission \citep[e.g.,][]{Luo2015,Ni2018,Pu2020,Ni2022}.
A few WLQs have also been found to vary between X-ray normal and X-ray weak states like
SDSS J$1350+2618$ but the established variability timescales are not as short 
\citep{Miniutti2012,Ni2020,Ni2022}.
The X-ray weakness and variability of the WLQs are naturally explained by the 
varying partial-covering absorption from the clumpy disk winds.
The weak line emission is probably due to
heavy shielding of the EUV/X-ray ionizing
photons from reaching the {C~{\sc iv}} BELR
by a thick inner accretion disk and/or its clumpy wind \citep[e.g.,][]{Luo2015,Ni2018,Ni2022}.
Apparently, not all such X-ray variable quasars are WLQs, and this might be 
related to the thickness of the inner accretion disk that mainly depends on the Eddington
ratio; i.e., WLQs are probably highly super-Eddington accreting with very thick disks.
Unfortunately, neither the SMBH masses nor the bolometric luminosities (e.g., see 
discussion in Section 4.3 of \citealt{Liu2021}) 
of super-Eddington accreting quasars can be reliably estimated, and thus much work is still
needed to understand the physical difference between WLQs and 
typical super-Eddington accreting quasars.

An alternative origin of the X-ray absorber is the 
thick inner accretion disk in these super-Eddington accreting AGNs,
as proposed in studies of the X-ray emission from WLQ samples \citep{Luo2015,Ni2018}.
The thick disk that shields the  {C~{\sc iv}} BELR might
be able to also block the X-ray emission
along the line of sight if the inclination angle is large. In this case, 
the observed fast \xray\ variability could be 
due to rotation of the thick inner disk that 
is somewhat azimuthally asymmetric \citep{Ni2020}. 
Compared to the clumpy disk wind as the absorber, the thick disk could provide heavier 
obscuration but its global covering factor is likely smaller. 
Better constraints on the frequency of X-ray variable
super-Eddington accreting AGNs and their X-ray weakness factors
will help to determine which absorber plays a more important role here.

Discovery of the extremely fast X-ray variability in SDSS J$1350+2618$ has considerable 
broader importance. It points a new direction 
to simply and elegantly identify high-velocity SMBH outflows
at high redshifts where high-quality X-ray spectra are not usually available.
Probably all such X-ray variable quasars possess extremely fast X-ray 
variability, but there have not been
any short-cadence X-ray monitoring observations of them, and 
only SDSS J$1350+2618$ has been caught in the act by the serendipitous
Chandra exposures. Understanding the SMBH growth and feedback processes of
super-Eddington accreting AGNs is especially important, as 
SMBH growth via super-Eddington accretion is likely
required to explain
the existence of massive SMBHs in the early universe \citep[e.g.,][]{Wu2015}, and 
super-Eddington accretion might be an important SMBH growth phase in general
in the high-redshift universe \citep[e.g.,][]{Netzer2007,Shen2012}.
X-ray monitoring observations of high-redshift quasars
might be able to reveal similar eclipse events, providing a new angle to 
probe SMBH growth and feedback at cosmic noon
($z\approx2$--3; the peak epoch of galaxy assembly and SMBH growth) and beyond.

\section{Summary}
We reported the extreme X-ray variability serendipitously detected in a $z=2.627$
radio-quiet type 1 quasar, SDSS J$135058.12+261855.2$, that has an estimated SMBH mass 
of $6.3\times 10^{9} M_\odot$.
It exhibited faint \xray\ emission
in 2015 August, with a rest-frame 2~keV flux density $\approx8.7$
times weaker compared to the expectation from its UV/optical
flux. In 2016 March (two months later in the quasar rest frame),
the second \chandra\ observation revealed that
the quasar recovered to an \xray\ nominal-strength state. The third observation in 2016 April
($47.2$ hours later in the rest frame) revealed that
the quasar had dimmed by a factor of $\approx 7.6$ in terms of its
\hbox{0.5--2 keV} flux. The dimming is associated with
spectral hardening, as the 2--7 keV flux dropped by only $17\%$.
The effective power-law photon index ($\Gamma_{\rm eff}$) of
the \xray\ spectrum changed from $2.3\pm0.4$ to $0.9_{-0.6}^{+0.7}$ (Table~1).
Such an extremely fast and \hbox{large-amplitude} X-ray variability
event has not been reported before in luminous quasars with such massive
SMBHs.
SDSS~J$1350+2618$ has a fairly typical quasar IR-to-UV SED (Figure~2) and a 
typical quasar rest-frame UV spectrum (Figure~4), and it 
does not show any
significant long-term variability in the IR and UV/optical bands (Figure~3).

The X-ray weak states of SDSS J$1350+2618$ are most naturally
explained by X-ray obscuration (Figure~5).
In the first observation, the line of sight to the X-ray emitting corona
was blocked by an absorber with
a neutral hydrogen column density of $\approx6.0\times 10^{23}~\rm cm^{-2}$
and a covering factor of $\approx93\%$.
The line of sight was cleared out during the second observation.
The extremely fast \xray\ dimming from the second to the third
observation is accounted for by an eclipse
event, where an X-ray absorber with a column density of
$\approx3.2\times10^{23}~\rm cm^{-2}$ moved across the line of sight
and fully ($\approx100\%$) covered the \xray\ corona.
Given the occultation time of $47.2$ hours,
and adopting a conservatively small size of $5r_{\rm g}$
for the \xray\
corona,
the velocity
of the absorber reaches a relativistic value of
$\approx 0.9c$, albeit with uncertainty from the
uncertain SMBH mass estimate.

SDSS J$1350+2618$ is likely accreting with a high
or even super-Eddington accretion rate, and it is closely connected to 
the super-Eddington accreting AGNs and WLQs displaying similar variability.
The high-velocity X-ray absorber is probably 
the dense gas clumps in the powerful accretion-disk wind. 
Such an energetic wind may eventually
evolve into a massive galactic-scale outflow,
expelling a large amount of gas and dust and
suppressing
host-galaxy star formation and further fueling of the SMBH.
Quasars like J$1350+2618$ are probably common in the high-redshift universe.
X-ray monitoring observations of these quasars might be able to reveal similar eclipse events,
providing a simple and elegant method to identify
high-velocity SMBH outflows and understand AGN feedback at cosmic noon
and beyond.

~\\

We thank the referee for the helpful comments.
H.L. and B.L. acknowledge financial support from
the National Natural Science Foundation of China
grant 11991053, China Manned Space Project grants NO. CMS-CSST-2021-A05
and NO. CMS-CSST-2021-A06.
H.L. acknowledges financial support from the program of China
Scholarships Council (No. 201906190104) for her visit in the
Pennsylvania State University. W.N.B. acknowledges support from the V.M. Willaman Endowment.

\clearpage
\begin{deluxetable*}{cccccccccccc}
\tablewidth{0pt}
\tabletypesize{\tiny}
\tablecaption{Chandra Observation Log and X-ray
Photometric Properties}
\tablehead{
\colhead{Obs.}                   &
\colhead{Observation}                   &
\colhead{Exposure}                   &
\colhead{Effective}                   &
\multicolumn{2}{c}{Net Counts}  &
\colhead{$\Gamma_{\rm eff}$}                   &
\multicolumn{2}{c}{Flux ($10^{-14}$ erg cm$^{-2}$ s$^{-1}$)}  &
\colhead{$f_{\rm 2~keV}$}                   &
\colhead{$\alpha_{\rm OX}$}                   &
\colhead{$\Delta\alpha_{\rm OX}$}                  \\
\cline{5-6}
\cline{8-9}
\colhead{ID}                   &
\colhead{Start Time}                   &
\colhead{Time (ks)}                   &
\colhead{Exp. (ks)}                   &
\colhead{0.5--2 keV}                   &
\colhead{2--7 keV}                   &
\colhead{ }                   &
\colhead{0.5--2 keV}                   &
\colhead{2--7 keV}                   &
\colhead{($10^{-32}$ erg cm$^{-2}$ s$^{-1}$ Hz$^{-1}$)}                   &
\colhead{ }                   &
\colhead{ }
}
\startdata
 17627  & 2015-08-28T16:29:50&55.5& 46.2&$\phantom{0}6.3_{-2.7}^{+ 4.0}$ &  $\phantom{0}6.2_{-3.2}^{+4.4}$ &$1.4_{-0.7}^{+ 0.9}$ & $0.19_{-0.08}^{+0.12}$    &        $0.39_{-0.20}^{+0.28}$  &$0.72_{-0.30}^{+0.46}$ &$-2.02$& $-0.36$
 \\
 17621  & 2016-03-27T05:25:18&62.8& 35.7 &$30.2_{-5.8}^{+7.0}$ & $12.8_{- 3.8}^{+5.2}$& $2.3_{-0.4}^{+ 0.4}$  & $1.14_{-0.22}^{+0.26}$   &    $0.70_{-0.21}^{+0.28}$&$7.71_{-1.48}^{+1.79}$ &$-1.62$& $\phantom{-}0.03$\\
 17222  & 2016-04-03T08:46:04&58.8& 48.4&$\phantom{0}6.1_{- 2.8}^{+4.1}$ & $11.3_{-4.4}^{+5.6}$& $0.9_{-0.6}^{+ 0.7}$ & $0.15_{-0.07}^{+0.10}$   & $0.58_{-0.22}^{+0.29}$ &$0.38_{-0.17}^{+0.26}$ &$-2.12$& $-0.47$  \\
\enddata
\tablecomments{The effective exposure is the exposure time
corrected for the effects of vignetting and CCD gaps.
All quoted errors are at a 68\% ($1\sigma$) confidence level. The errors of
the net counts were propagated from the errors of the source and background counts.}
\end{deluxetable*}

\end{document}